\documentclass[12pt]{iopart}
\usepackage{graphicx}
\usepackage{color}
\usepackage{subfigure}

\usepackage{soul}

\begin{document}
\title
{AQPDCITY Dataset: Picture-Based PM Monitoring in the Urban Area of Big Cities}
\author{Yonghui Zhang$^{1-4}$, Ke Gu$^{1-4}$}
\address{
$^1$Engineering Research Center of Intelligent Perception and Autonomous Control, \tiny\textcolor{white}{A}\small Ministry of Education \\
$^2$Beijing Key Laboratory of Computational Intelligence and Intelligent System \\
$^3$Beijing Artificial Intelligence Institute \\
$^4$Faculty of Information Technology, Beijing University of Technology, China\\}
\ead{guke.doctor@gmail.com}
\vspace{10pt}
\begin{indented}
\item[]November 2019
\end{indented}

\begin{abstract}
Since Particulate Matters (PMs) are closely related to people's living and health, it has become one of the most important indicator of air quality monitoring around the world. But the existing sensor-based methods for PM monitoring have remarkable disadvantages, such as low-density monitoring stations and high-requirement monitoring conditions. It is highly desired to devise a method that can obtain the PM concentration at any location for the following air quality control in time.
The prior works indicate that the PM concentration can be monitored by using ubiquitous photos. To further investigate such issue, we gathered 1,500 photos in big cities to establish a new AQPDCITY dataset. Experiments conducted to check nine state-of-the-art methods on this dataset show that the performance of those above methods perform poorly in the AQPDCITY dataset.
\end{abstract}

\section{Introduction}
With the repaid development of industrialization and urbanization, the emission of air pollutants increase gradually, especially Particulate Matters (PMs). As we all know, PMs do serious harm to human health and and air quality. For example, PMs are one of the most critical reasons in casing lung cancer \cite{1}. Besides, it is easy to case serious damages in skin and mucous membrane of someone who long-timely exposed in high-concentration PM, and thus makes the sicken probability of skin cancer is much higher than that in the 1990s \cite{2}-\cite{5}. PMs also cause serious environment problems, especially atmospheric environment \cite{6e}-\cite{6f}.

\begin{figure}[!t]
\vspace{0.125cm}
\centering
\subfigure[]{\label{fig1:subfig:a}\includegraphics[width=3.5cm]{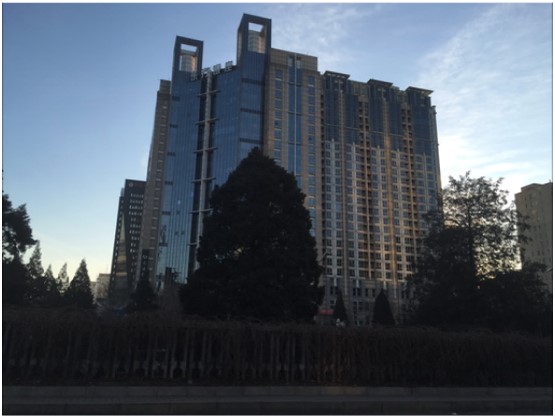}}\hspace{0.1cm}
\subfigure[]{\label{fig2:subfig:b}\includegraphics[width=3.5cm]{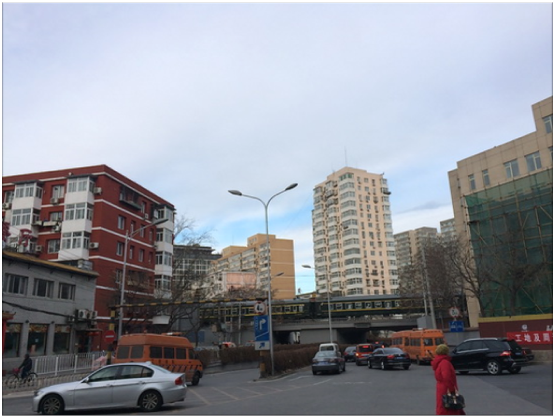}}\hspace{0.1cm}
\subfigure[]{\label{fig3:subfig:c}\includegraphics[width=3.5cm]{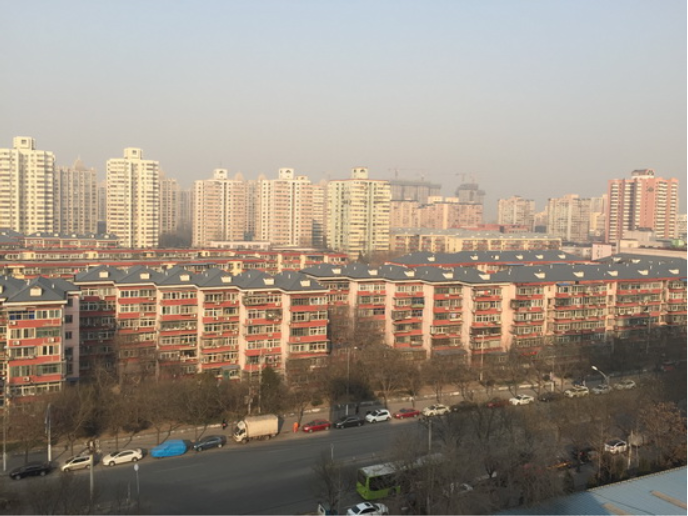}}\\
\subfigure[]{\label{fig3:subfig:d}\includegraphics[width=3.5cm]{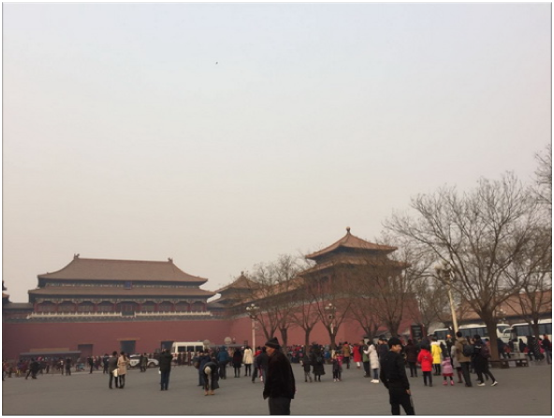}}\hspace{0.1cm}
\subfigure[]{\label{fig3:subfig:e}\includegraphics[width=3.5cm]{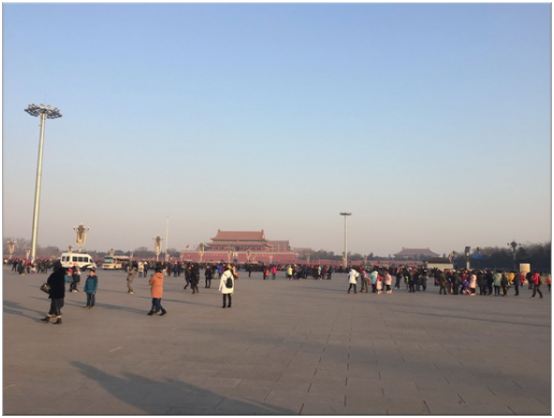}}\hspace{0.1cm}
\subfigure[]{\label{fig3:subfig:f}\includegraphics[width=3.5cm]{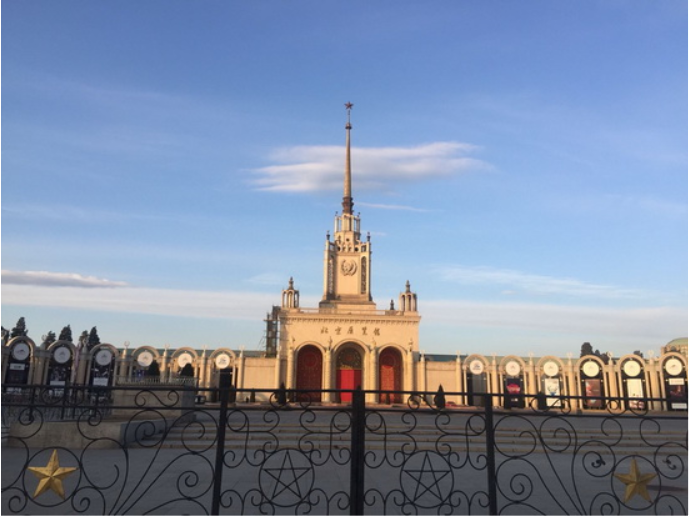}}\\
\subfigure[]{\label{fig3:subfig:d}\includegraphics[width=3.5cm]{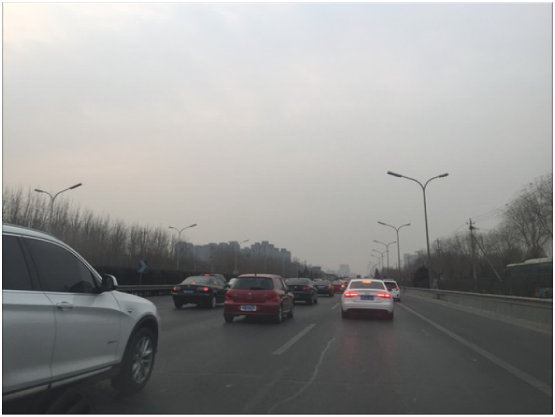}}\hspace{0.1cm}
\subfigure[]{\label{fig3:subfig:e}\includegraphics[width=3.5cm]{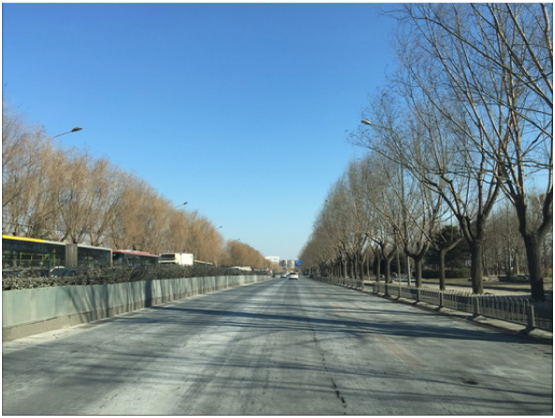}}\hspace{0.1cm}
\subfigure[]{\label{fig3:subfig:f}\includegraphics[width=3.5cm]{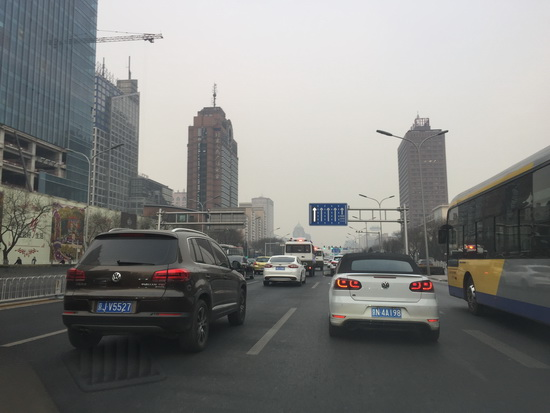}}\\
\vspace{0.1cm}
\small \textbf{Figure 1.} Example photos with different scenes in the AQPDBJUT dataset: (a)-(c) Buildings; (d)-(e) Scenic spots; (f)-(g) Roads.
\label{fig:1}
\end{figure}

As a data source for air quality prevention and control, the existing sensor-based stations are incomplete for full PM concentration monitoring. To promote the innovation and development of PM monitoring, Gu \emph{et al.} have already made several achievements by conducting a long-term research \cite{3}-\cite{9f}.

A high-density real-time PM monitoring map can provide more effective advice for PM prevention and control. In order to obtain a high-density PM monitoring map, during the last three years, we have collected 1,500 photos in big cities to establish a novel dataset dubbed as AQPDCITY. These photos were captured in different places and different times, thus having a wide coverage of photo scenes. Experiments show that the performances of state-of-the-art relevant methods are not ideal in PM monitoring.

\section{Dataset}
The AQPDCITY dataset contains 1,500 photos which were captured at the various locations and seasons and under the different weather situations during the past three years. The equipments used are smartphone and a single-lens reflex camera, etc., as illustrated in Fig. \ref{fig:1}. This dataset includes diversified scenes of squares, temples, lakes, streets, overpasses, buildings, cars, parks and so on, which makes it better reflect the distribution of natural landscape in the city. All the photos in the AQPDCITY dataset have a wide range of resolutions from 500$\times$261 to 978$\times$550. For each photo, we obtain its PM value by searching the historical record of hourly PM concentration from the China National Environmental Monitoring Centre (CNEMC). The PM concentration has a wide range from 0 to 634.

\begin{figure}
\vspace{0.125cm}
\centering{\small \textbf{Table 1.} Comparison of nine state-of-the-art models for \\ PM$_{2.5}$ monitoring on the AQPDCITY dataset.}\\
\centering
\vspace{0.3cm}
\label{fig2:subfig:b}\includegraphics[width=9cm]{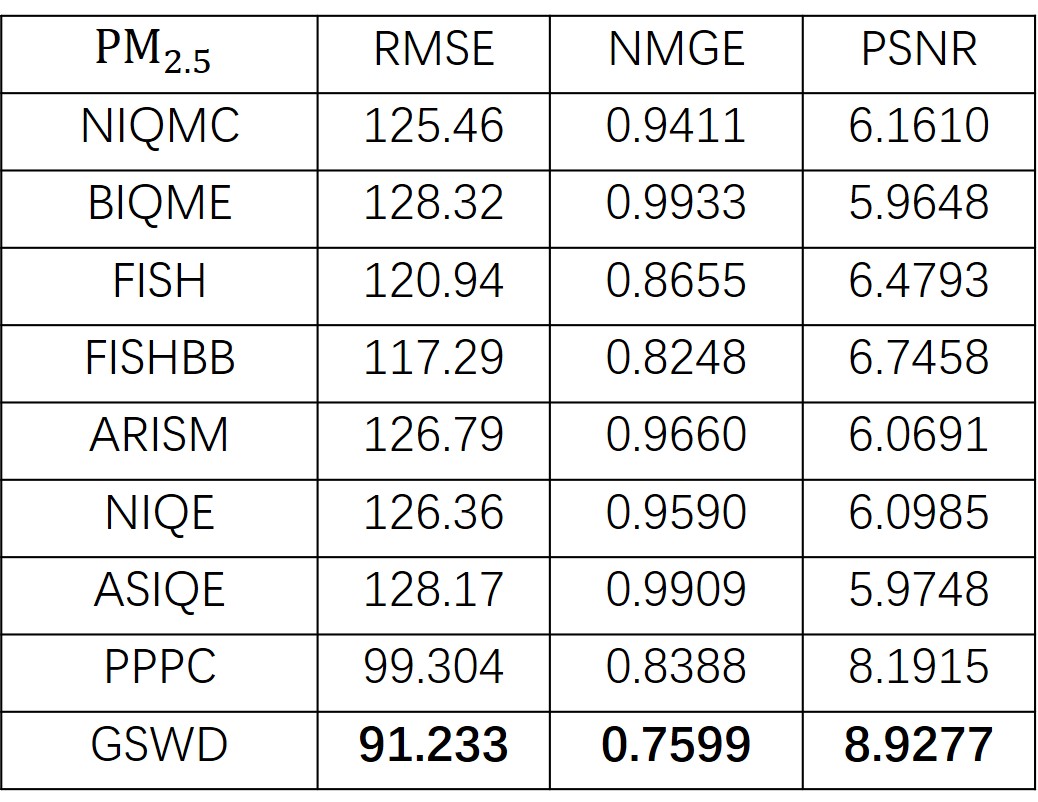}\\
\label{tab:1}
\end{figure}

\section{Experiment}
We compute the performance of nine state-of-the-art models on the AQPDCITY datasets, by using three indicators including the Root Mean Square Error (RMSE), the Normalized Mean Gross Error (NMGE) and the Peak Signal to Noise Ratio (PSNR). Nine state-of-the-art photo-based methods include NIQMC \cite{35}, BIQME \cite{6g}, FISH \cite{36}, FISHBB \cite{36}, ARISM \cite{6h}, NIQE \cite{6k}, ASIQE \cite{6j}, PPPC \cite{6d}, and GSWD \cite{4}, as highlighted in the table.
The newly proposed GSWD has obtained the optimal performance.

\section{Conclusion}
In this paper we have established a new AQPDCITY dataset which consists of 1,500 photos gathered in big cities. Those photos includes various scenes, such as squares, temples, lakes, streets, overpasses, buildings, cars, parks and so on. Experiments show the performance of state-of-the-art methods perform poorly in that dataset. The control-oriented photo PM monitoring method is in the stage of rapid development, a new dataset can promote the improvement of the PM monitoring method.

\end{document}